\documentstyle[prl,aps,multicol]{revtex}  
\input{psfig}
\begin{document}
\draft
\title{
Random Matrix Theory of a Chaotic Andreev Quantum Dot}
\author{Alexander Altland and Martin R. Zirnbauer}
\address{Institut f\"{u}r Theoretische Physik, 
Universit\"{a}t zu K\"{o}ln, Z\"{u}lpicher Str. 77, 
50937 K\"{o}ln, Germany}
\date{February 20, 1996}
\maketitle
\begin{abstract}
A new universality class distinct from the standard
Wigner-Dyson ones is identified. This class is realized
by putting a metallic quantum dot in contact with a
superconductor, while applying a magnetic field
so as to make the pairing field effectively vanish on
average.  A random-matrix description of the spectral
and transport properties of such a quantum dot is
proposed. The weak-localization correction to the
tunnel conductance is nonzero and results from the
depletion of the density of states due to the coupling
with the superconductor.  Semiclassically, the
depletion is caused by a singular mode of
phase-coherent long-range propagation of particles and
holes. 
\end{abstract}

\begin{multicols}{2}

Wigner-Dyson level statistics is found in physical
systems as diverse as highly excited molecules, atoms
and nuclei, mesoscopic systems in the ballistic or
diffusive regime, and chaotic Hamiltonian systems such
as the stadium or Sinai's billiard. The reason for the
ubiquity and universality of Wigner-Dyson statistics is
the relation of the Gaussian ensembles\cite{mehta} to
attractive fixed points of the renormalization group
flow for an effective field theory (nonlinear $\sigma$
model)\cite{efetov}.  Depending on whether time
reversal and/or spin rotation invariance is broken or
not, the relevant ensemble has orthogonal, unitary, or
symplectic symmetry.

Although the Wigner-Dyson ensembles are the generic
ones, new universality classes may arise when
additional symmetries or constraints are imposed. One
example of this is provided by systems with A-B
sublattice structure or chiral symmetry\cite{gade}.
Another example, presented in this letter, are metallic
systems in contact with a superconductor. Our
considerations were inspired in part by
work\cite{oppermann} on Anderson localization of normal
excitations in a dirty superconductor or a
superconducting glass. 

\narrowtext
\begin{figure}
\centerline{\psfig{figure=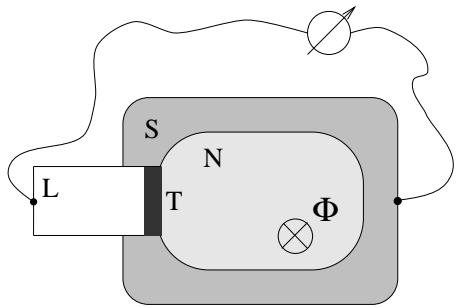,height=4.0cm}}
\vspace{0.5cm}
\caption{Metallic quantum dot (N) in contact with a
superconductor (S). A tunnel barrier (T) separates
the dot from a normal-metal lead (L) attached for the
purpose of making current and voltage measurements.}
\end{figure}

The system we have in mind is depicted in Fig.~1: a
metallic (normal-conducting) quantum dot (N) with the
shape of, say, a stadium billiard, is surrounded by a
superconductor (S) and contacted by a normal-metal lead
(L) via a tunnel barrier (T). A Schottky (or potential)
barrier forms at the NS-interface. The quantum dot is
pierced by a weak magnetic flux of the order of one or
several flux quanta. The temperature is so low that the
phase-coherence length exceeds the system size by far.
It will help our argument if the quantum dot is not
perfectly shaped but contains some defects and/or
impurities.

\narrowtext
\begin{figure}
\centerline{\psfig{figure=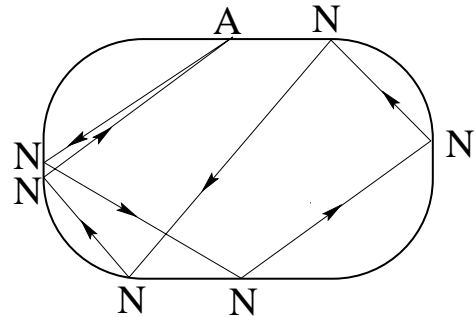,height=4.0cm}}
\vspace{0.5cm}
\caption{(Almost) periodic orbit involving several normal 
reflections (N) and one Andreev reflection (A). An
electron runs through the closed loop twice. Before
reaching A it is a particle and afterwards a hole.}
\end{figure}

The unique feature that distinguishes our quantum dot
from conventional mesoscopic systems is the process of
Andreev reflection\cite{andreev}, by which an
electron incident on the NS-interface is retroreflected
as a hole, and vice versa.  The drastic consequences of
this process for the thermodynamics and the transport
of the quantum dot can be anticipated from a
semiclassical argument. Consider the periodic
orbit of Fig.~2: an electron starts out from point A,
undergoes several normal reflections off the Schottky
barrier and eventually returns to A, with the final
velocity being roughly the negative of the initial one.
At the point of return, the electron gets converted
into a hole by Andreev reflection.  Because the
magnetic field is too weak to cause any significant
bending of classical trajectories, the hole then simply
tracks down the path laid out by the electron. The
charge of the hole is opposite to that of the electron,
so the magnetic phase accumulated along the total path
is zero.  Moreover, since a hole is not just the charge
conjugate but also the time reverse of an electron, the
dynamical phases cancel, too, provided that electron
and hole are at the same energy (the Fermi energy).
The two Andreev reflections add up to a total phase
shift of $\pi$.  Thus, the periodic orbit of Fig.~2
contributes to the periodic-orbit sum for the density
of states {\it with the negative sign}. Orbits of this
type reduce the mean density
of states (``Weyl term'') and are expected to put our
``Andreev quantum dot'' in a universality class
distinct from the standard Wigner-Dyson ones. We shall
identify this universality class in the sequel.

In a microscopic mean-field treatment, we would start 
from the Bogoliubov-deGennes (BdG) Hamiltonian
        \[
        {\cal H} = \pmatrix{       
        H_0       &\Delta       \cr
        \Delta^*  &-H_0^*         \cr}    
        \] 
where $H_0 = ({\bf p}-e{\bf A})^2/2m + V({\bf x}) -
\mu$ is a Hamiltonian for ``particles'', $-H_0^*$ is the
corresponding Hamiltonian for ``holes'', and the pairing
field $\Delta({\bf x})$, whose magnitude rises from
zero inside the quantum dot to a nonzero value in the
superconductor, converts particles into holes. The
potential $V({\bf x})$ includes the Schottky barrier.
Energy is measured relative to the chemical potential 
$\mu$.

Universality classes are characterized by their
symmetries.  Notice therefore that ${\cal H}$ obeys
the relation
        \begin{equation}
        {\cal H} = - {\cal C} {\cal H}^{\rm T} 
        {\cal C}^{-1}, \quad {\cal C} = \pmatrix{
        0       &{\bf 1}        \cr
        -{\bf 1}&0              \cr} .
        \label{CT}
        \end{equation} 
This symmetry originates from the electron's being a
spin-1/2 fermion and from spin-rotation invariance
\cite{footnoteA}.  The transformation ${\cal H} \mapsto
{\cal C} {\cal H}^{\rm T} {\cal C}^{-1}$ will be called
${\cal CT}$-conjugation as it combines time reversal
(${\cal H} \mapsto {\cal H}^* = {\cal H}^{\rm T}$) with
a kind of charge conjugation (${\cal H} \mapsto {\cal C
H C}^{-1}$). 

By design, the classical motion of particles and holes
in the billiard-shaped dot is chaotic and fills the
available phase space ergodically. Now observe that
every time a particle or hole is Andreev-reflected from
the NS-interface, its wavefunction acquires an extra
phase determined by the superconducting order
parameter.  In this context it is important that the
applied magnetic field is screened by a supercurrent
circulating along the NS-interface inside the
superconductor.  The supercurrent flow, in turn, is
concomitant with a spatial variation of the phase
$\phi$ of the order parameter\cite{footnoteB}.  As a
result of this and the chaotic dynamics, the extra
phase picked up during Andreev reflection varies
randomly along a typical semiclassical trajectory.  So
everything is quite random, and we expect some kind of
random-matrix theory to apply. The question is now:
what random-matrix theory?

Because the presence of the magnetic field makes the pairing 
field experienced by particles and holes vanish on average,
$\Delta$ can be modelled by a stochastic variable with zero 
mean.  Moreover, since the system has been
designed to be chaotic, there exist no integrals of
motion except for energy.  The only symmetry (apart
from hermiticity of the Hamiltonian) of relevance for
the long-time\cite{footnote1} or ergodic limit we shall
consider, is the ${\cal CT}$-oddness (\ref{CT}).
Experience with similar problems then tells us that we
can model the ergodic limit by a Gaussian, or
maximum-entropy, ensemble with probability density
$\exp (-{\rm Tr} {\cal H}^2/2v^2) d{\cal H}$ subject to
the constraint (\ref{CT}). This implies that, in any
orthonormal basis of states $\psi_a$ $(a=1,{\bar 1},
...,N,{\bar N})$ with ${\cal CT}$-conjugate basis
$\psi_{\bar a} = {\cal C}\psi_a^*$, the variances of
the random Hamiltonian matrix elements are given by the
correlation law
	\begin{equation} \langle {\cal H}_{ab}
	{\cal H}_{cd} \rangle = v^2 (\delta_{ad}
	\delta_{bc} - \delta_{a\bar c}\delta_{b\bar d}).  
	\label{correlator}
	\end{equation} 
To complete the definition of our random-matrix model,
we add to ${\cal H}$ a term $-i\Gamma$ which accounts
for the coupling to the normal-metal lead and will be
specified later.

As a first step, let us close off the contact with the
lead $(\Gamma = 0)$. What can we say about the spectral
statistics, the central characteristic of the ergodic
{\it isolated} quantum dot? If $\psi_k$ is an
eigenstate of ${\cal H}$ with eigenvalue $+E_k$ then,
by (\ref{CT}), so is ${\cal C}\psi_k^*$ with eigenvalue
$-E_k$.  Thus, there is an exact pairing between
positive and negative eigenvalues. By diagonalizing
${\cal H}$ and computing the Jacobian of the
transformation to diagonal form, we obtain the 
(unnormalized) joint probability density of the 
positive eigenvalues $E_k$ $(k=1,...,N)$:
        \begin{equation} 
        P(E_1,...,E_N) = \prod_{i<j}
        (E_i^2-E_j^2)^2 \prod_{k=1}^N E_k^2 \ 
        e^{-E_k^2/2v^2} , 
        \label{ensemble} 
        \end{equation} 
which is manifestly invariant under $E_k \to 
-E_k$\cite{nagao}.

To calculate the spectral statistics, it is convenient
to view (\ref{ensemble}) as a Gaussian Unitary Ensemble
(GUE) of $2N$ levels $E_1, E_{\bar 1}, ..., E_N,
E_{\bar N}$ with the mirror constraint $E_{\bar k} = -
E_k$.  The correlation functions of the GUE are known
to coincide with those of a one-dimensional gas of free
fermions in the large-$N$ limit\cite{sutherland}. Now,
when a fermion (i.e. an energy level) gets close to $E
= 0$, so does its mirror image.  Because Fermi
statistics makes the wavefunction vanish as two
fermions approach, the constraint $E_{\bar k} = -E_k$
amounts to hard wall boundary conditions at $E = 0$.
Hence, we can compute the eigenvalue density and its
correlations for (\ref{ensemble}) as the {\it particle
density and its correlations for a free Fermi gas with
a hard wall at the origin}. In this way we obtain
	\begin{equation} \rho(E) = \langle
	{\rm Tr} \delta(E-{\cal H})\rangle = 
	1/\delta - \sin (2\pi E/\delta) / 2\pi E ,
	\label{density} 
	\end{equation} 
where $\delta$ is the level spacing for $E \gg \delta$.
We see that the coupling to the superconductor depletes
the mean density of states and makes it vanish
quadratically at $E = 0$\cite{footnote2}. The states
pushed away from $E = 0$ cause density oscillations,
which ebb off as $1/E$. Note that the result
(\ref{density}) applies when both $E$ and $\delta$ are
much smaller than the characteristic energy uncertainty
set by the frequency of Andreev reflection.

To understand better the mechanism of depletion, we
turn to diagrammatic perturbation theory. Let $G :=
\langle (E+i\varepsilon-{\cal H})^{-1}\rangle$ denote  
the ensemble average of the Gorkov Green's function.
We expand it in a geometric series with respect to
${\cal H}$ as usual.  To do the ensemble average, we
distinguish between two types of contraction,
$\Pi_{ac,bd}^{\rm D} = v^2 \delta_{ad}\delta_{bc}$ and
$\Pi_{ac,bd}^{\rm C} = -v^2 \delta_{a\bar c}
\delta_{b\bar d}$, corresponding to the first and
second term in the basic law (\ref{correlator}).
Making this distinction is useful for organizing the
perturbation series, since $\Pi^{\rm D}$ causes pure
GUE behavior whereas $\Pi^{\rm C}$ generates the
corrections to the GUE.  By summing all nested
$\Pi^{\rm D}$ self-energy graphs, we get Pastur's
equation, $G = (E+i\varepsilon - v^2 {\rm Tr}G)^{-1}$,
which is exact for $E \gg \delta$ and $N\to\infty$. The
solution, $G^0$, of this equation yields Wigner's
semicircle law for the density of states: $-{\rm Im
Tr}G^0/\pi = \sqrt{2N-(E/v)^2}/\pi v$.  According to
(\ref{density}), corrections to this result, which is
stationary and equal to $\sqrt{2N}/\pi v =: 1/\delta$
up to uninteresting terms of order $1/N$, should appear
as we approach zero energy.  It turns out that these
arise from summing a geometric series of ladder graphs
built solely from $\Pi^{\rm C}$-contractions. The
ladder sum, $C_{ac,bd}$, satisfies Dyson's equation $C=
C^0 + C^0 \Pi^{\rm C} C$ with $C_{ac,bd}^0 =
\delta_{ab} \delta_{cd} G_{aa}^0 G_{cc}^0$. Its
solution $C = C^0({\bf 1} -\Pi^{\rm C}C^0)^{-1}$ is
{\it singular} at $E = 0$:
        \[
        C = C^0 + C^0 \left( 
        \Pi^{\rm C} 2iN\delta/\pi E \right) C^0 + 
        {\cal O}(E/N)^0.
        \] 
By evaluating the graph shown in Fig.~3 we get ${\rm
Tr}G/\pi = -i/\delta - 1/2\pi E + ...$, which are the
leading terms in a $1/E$ expansion. (There is a
renormalization by a factor of 1/2 coming from the
possibility of connecting the external legs in Fig.~3
by a nonsingular $\Pi^{\rm D}$ ladder.) This
perturbative result is to be compared with the exact
formula ${\rm Tr}G/\pi = -i/\delta - [1-\exp(2\pi
iE/\delta)]/2\pi E$ reconstructed from (\ref{density})
by causality.  We see that diagrammatic perturbation
theory properly reproduces the smooth part of the $1/E$
correction.  (The oscillatory term is nonanalytic in
the expansion parameter $1/E$ and cannot be recovered
by the perturbative summation of graphs.)

What is the semiclassical meaning of the mode $C$?  Our
diagrammatic analysis suggests an interpretation as a
mode of phase-coherent propagation of a particle-hole
pair. To gain further insight, recall the periodic
orbit shown in Fig.~2. In diagrammatic language, this
orbit corresponds to connecting the Green's function
lines in Fig.~3 by just a single $\Pi^{\rm
C}$-contraction, and thus is the simplest semiclassical
building block of the $C$-mode.  More generally, an
arbitrary number of Andreev reflections may be inserted
into the loop. The only condition is that the {\it
particle-hole character of the states during the first
and second traversal of the loop be ${\cal
CT}$-conjugate to each other}.  Because ${\cal
CT}$-conjugate states carry opposite charge and the
loop is traversed twice {\it in the same direction},
the $C$-mode is ignorant of a weak magnetic
field\cite{footnoteC}.  Note, however, that the
$C$-mode is sensitive to energy $E$ (or voltage), which
breaks the phase relation between particles and holes.

\narrowtext
\begin{figure}
\centerline{\psfig{figure=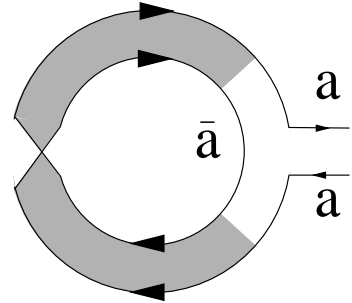,height=4.0cm}}
\vspace{0.5cm}
\caption{Diagram contributing to the average Green's
function in order $1/E$. The shaded region symbolizes
the $C$-mode.}
\end{figure}

With a solid understanding of the spectral properties
in hand, we finally open up the tunnel barrier and turn
to the prime experimental observable, the conductance 
$g$. Our treatment will be based on the linear response
formula $g=4(e^2/h){\rm Tr}{S^\dagger}_{\rm ph}S_{\rm hp}
$\cite{takane} where $S_{\rm hp}$ is the part of the
scattering matrix $S$ that maps incoming electrons onto
outgoing holes.  We suppress elastic phase shifts and  
parameterize the $S$-matrix by
        $
        S = {\bf 1} - 2i{\tilde W}(i\varepsilon+i\Gamma
        -{\cal H})^{-1} W
        $ 
\cite{iwz} at $E = 0$.  Here $W$ is a matrix coupling $2M$
channels (particles and holes) in the lead with $2N$
levels in the quantum dot, $\tilde W$ is its adjoint,
and $\Gamma = W{\tilde W}$.  $W$ is diagonal in
particle-hole space. For a simple model, we take
${\tilde W}W$ to be a multiple of the identity in
channel space: ${\tilde W}W = \gamma{\bf 1}$. Then
$\Gamma = \gamma P$ where $P$ is a rank-$2M$ projector
in level space. We assume $N \gg M$.

We begin by discussing the limit of an open quantum dot
$(M\gg 1)$ where all structure in the density of states
is washed out by the large level width. Let $T$ denote
the probability for an electron in an incoming channel
to be transmitted through the tunnel barrier, in this
very limit. For our simple model, $T = 4\lambda\gamma/
(\lambda +\gamma)^2$ with $\lambda^2 = 2Nv^2$
\cite{iwz}.  Having entered the quantum dot, the
electron spends a long time there, undergoing many
Andreev reflections, and finally returns to the lead
either as a particle or as a hole, with equal
probabilities. Hence $\langle {\rm Tr}{S^\dagger}_{\rm
ph}S_{\rm hp}\rangle = MT/2$, which we refer to as the
``classical'' value. To confirm this result
diagrammatically, one has to sum a geometric series of
$\Pi^{\rm D}$ ladder graphs, producing the
diffuson\cite{footnote3}.  On approaching the opposite
(closed) limit, we expect a reduction relative to the
classical value.  The reason is simply this: the
density of states in the isolated dot vanishes at $E =
0$, so that an electron trying to enter cannot find any
state to go to and is immediately rejected into the
lead.  This effect should announce itself as a negative
${\cal O}(M^0)$ correction (``weak localization'').
Such a correction in fact exists and is due to the
graph shown in Fig.~4 featuring two diffusons and one
$C$--mode. Evaluation of this graph leads to
        \begin{equation}
        \langle g \rangle = (MT - 1 + T) \times 2e^2/h
        + {\cal O}(1/MT) .
        \label{conductance}
        \end{equation} 
Note that the weak localization correction in 
the corresponding N-system vanishes. A result 
similar to (\ref{conductance}) was recently 
found in a related context\cite{brouwer}. 

\begin{figure}
\centerline{\psfig{figure=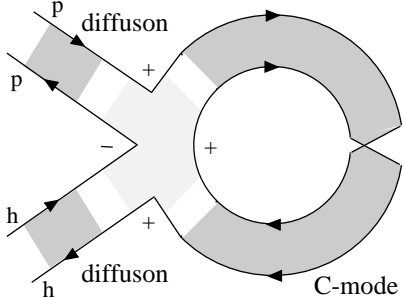,height=4.0cm}}
\vspace{0.5cm}
\caption{Diagram contributing to the average conductance
in order $M^0$. The shading of the vertex center
indicates unitarity-conserving (Hikami box) corrections.}
\end{figure}

Although the calculation of the graph of Fig.~4 is
somewhat technical, the result can easily be checked
and interpreted. Consider first the limit of many
weakly coupled channels ($T\to 0$, $MT$ fixed).  In
this limit the absolute square of the Green's function
becomes independent of the level and particle-hole
indices, on average\cite{mrznpa}. By conservation of
probability, the conductance is then fully determined
by a one-point function: $\langle g \rangle = -(4e^2/h)
\bar\gamma {\rm Im} \langle {\rm Tr} (i\bar\gamma-{\cal 
H})^{-1}\rangle$ where $\bar\gamma = \gamma M/N$ is the 
mean level width.  Analytic continuation of our result 
for the average Green's function to $E = i\bar\gamma$ 
gives $\langle g\rangle = (MT-1+e^{-MT})\times 2e^2/h$, 
in agreement with (\ref{conductance}).

To understand the other limit of strongly coupled
channels $(T=1)$ it is easiest to argue directly at the
$S$--matrix level. By its definition as a ``sticking
probability'', $T = 1 - |\langle S_{cc}\rangle |^2
+{\cal O}(1/MT)$. Putting $T = 1$ therefore
amounts to assuming a fully random $S$-matrix that
vanishes on average: $\langle S \rangle = 0$. The
symmetries of $S$ are the same as those of $\exp i{\cal
H}$, so that from (\ref{CT}) we deduce $S = {\cal C}
{S^{-1}}^{\rm T} {\cal C}^{-1}$, which is the defining
equation of the symplectic group ${\rm Sp}(2M,{\bf
C})$.  (${\cal C}$ is now meant to operate in channel
space.)  Unitarity fixes a subgroup ${\rm Sp}(2M)$. We
are thus led to postulate an $S$-matrix ensemble with
probability distribution $\langle\langle\bullet\rangle
\rangle$ given by the Haar (or invariant) measure of 
${\rm Sp}(2M)$. This ensemble is Dyson's Circular 
Unitary Ensemble\cite{mehta} adapted to fit our particle-hole
symmetric situation.  By elementary group theory, 
$\langle\langle S_{ab}\rangle\rangle = 0$ and $\langle
\langle S_{ab}^{\vphantom{*}} S_{cd}^* \rangle\rangle = 
\delta_{ac}\delta_{bd}/2M$, so $\langle\langle g \rangle 
\rangle = 2Me^2/h$, with no correction of order $M^0$.

In conclusion, we mention that the singular modes, the
diffuson and the $C$-mode, translate into low-energy
degrees of freedom of a corresponding nonlinear $\sigma$ 
model. They determine the attractive (metallic, or free)
renormalization group fixed point of this field theory.
Ultimately, the existence of such singular modes, which
are forgetful of microscopic detail and saturate the
long-time and long-distance physics, is the reason why 
we predict with confidence that the ergodic limit of 
the chaotic Andreev quantum dot is universal and our
random-matrix theory applies.

\end{multicols}

\begin{thebibliography}{99}
\bibitem{footnoteB} Note that the choice of gauge $\phi = 
{\rm const}$ is unnatural here as it forces $\int {\bf A}d{\bf x}$
across the lead to be very large.
\bibitem{footnote1}
Here ``long'' means times greater than the mean time $t_A$
spent between successive Andreev reflections.  In order for
the ergodic limit to be nonempty, we require $t_A \gg \hbar
/\delta$ where $\delta$ is defined in the text below 
Eq.~(\ref{density}).
\bibitem{footnote2} For lack of space the correlations are not
recorded here.
\bibitem{footnoteC} By this feature, the $C$-mode can be identified
as a close relative of the usual (normal-metal) diffuson. It here
appears as a separate mode because of our use of the BdG-formalism.
\bibitem{footnote3} The diffuson is identical to the normal-metal 
diffuson except that in our case the opposite segments of the 
ladder are either two BdG-particles or two BdG-holes.
\end{thebibliography}
\end{document}